# DC-powered $Fe^{3+}$:sapphire Maser and its Sensitivity to Ultraviolet Light

Mark Oxborrow[(1)], Pierre-Yves Bourgeois[(2)], Yann Kersalé [(2)] and Vincent Giordano[(2)]

*[(1)]National Physical Laboratory*
*Hampton Road, Teddington, Middlesex TW11 0LW, United Kingdom*
*Email: mo@npl.co.uk*

*[(2)]Institut FEMTO-ST, UMR 6174 CNRS-Université de Franche-Comté*
*32 av. de l'Observatoire, 25044 Besançon Cedex, France*
*Email: vincent.giordano@femto-st.fr*

## INTRODUCTION

As realized early on [1] [2], the zero-field $Fe^{3+}$-doped-sapphire maser variant of the whispering-gallery-mode cryogenic sapphire oscillator (CSO) exhibits several alluring features: Its output is many orders of magnitude brighter than that of an active hydrogen maser and thus far less degraded by spontaneous-emission (Schawlow-Townes) and/or receiving-amplifier noise. Its oscillator loop is confined to a piece of mono-crystalline rock bolted into a metal can. Its quiet amplification combined with high resonator $Q$ provide the ingredients for exceptionally low phase noise [3]. We here concentrate on novelties addressing the fundamental conundrums and technical challenges that impede progress:

(A) *Roasting:* The "mase-ability" of sapphire depends significantly on the chemical conditions under which it is grown and heat-treated. Beyond merely confirming previous work [4], we provide some fresh details and nuances.

(B) *Simplification:* This paper obviates the need for a Ka-band synthesizer: it describes how a 31.3 GHz loop oscillator, operating on the preferred WG pump mode, incorporating Pound locking, was built from low-cost components.

(C) *"Dark Matter"*: A Siegman-level [5] analysis of the experimental data determines the substitutional concentration of $Fe^{3+}$ in HEMEX to be less than a part per billion prior to roasting and up to a few hundred ppb afterwards [4]. Chemical assays, using different techniques (incl. glow discharge mass spectra spectroscopy and neutron activation analysis) consistently indicate, however, that HEMEX contains iron at concentrations of a few parts per million. Drawing from several forgotten-about/under-appreciated papers, this substantial discrepancy is addressed.

(D) *Excitons:* Towards providing a new means of controlling the $Fe^{3+}$:sapph. system, a cryogenic sapphire ring was illuminated, whilst masing, with UV light at wavelengths corresponding to known electronic and charge-transfer (thus valence-altering) transitions. Preliminary experiments are reported.

## SAPPHIRE ROASTING TRIALS

In contrast to Czochralski growth, the 2050°C graphite heaters within a HEM growth furnace [6] produce a reducing atmosphere that induces oxygen vacancies and lowers the valence of ionic impurities. Roasting, *i.e.* annealing at high temperature in air, oxidizes $Fe^{2+}$ ions inside this sapphire (back) to $Fe^{3+}$, though it is hypothesized here that the complete chemistry, as it affects ESR strengths, is more complicated than this process alone: the concentration and/or valencies of (i) defects (specifically F centers), (ii) individual impurities (*e.g.,* Ti, Mo), potentially connected to $Fe^{2+}/Fe^{3+}$ through charge-compensation, and (iii) $Fe^{3+}$:$Fe^{3+}$ clusters, amid a zoo of other combinations, will also be modified [7, 8].

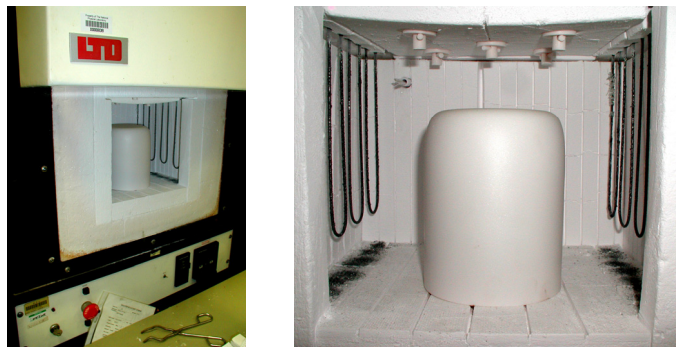

Fig. 1: Lenton chamber furnace for sapphire annealing

Two HEMEX sapphire resonators, *viz*. Léonard and Basile (see Table 1), were received from FEMTO-ST. It was noticed that both contained optical striae. On mounting into a copper can of standard design, bolted directly to the second stage of a Cryomech PT405 pulse-tube cooler, both supported a $WGH_{17}$ doublet near 12.03 GHz, with *Qs* of around 60 million at 3 K, exhibiting slight bistabilities. On pumping on all candidate WG modes near 31.3 GHz with a Wiltron 6742B synthesizer boosted by an Agilent 83050A [2-50 GHz, 100mW] amplifier, neither sapphire ring could be gotten to mase. Both were subsequently subjected to roasting.

This was accomplished with a Lenton chamber furnace [9]; 200×200×200 mm sample space, maximum operating temperature 1700° C; $MoSi_2$ heating elements; programmable Eurotherm controller with Pt6Rh/Pt30Rh (type B) thermocouple. Each sapphire ring was cleaned (incl. soak in Piranha bath then DI rinse) and placed onto a 10-mm diameter alumina pedestal inside the chamber; a 100-mm diameter alumina beaker was thereupon placed over the ring to protect the latter from debris shedded by the furnace's liner and heaters. After roasting, Léonard contained a feint, internal milky "cloud" near its cylindrical surface, presumably oxidized impurities. Roasting did not noticeably alter the *Q* of either sapphire ring. Both mased on driving (down to 0 dBm) a variety of $WG_{XX}$ pump modes near 31.3 GHz. Léonard exhibited bistable masing (dependent on coupling), the two modes of its $WGH_{17}$ doublet separated by 23 kHz.

| Table 1 | Ø [mm] | height [mm] | annealing schedule | (pump) $WG_{XX}$ freq. [GHz] | signal ($WGH_{17}$) freq. [GHz] | mode line-width [Hz] | maser output power |
|---|---|---|---|---|---|---|---|
| Léonard | 50.017 | 30.018 | 16 hour @ 1600° C; 200° C/hr ramp up/down. | 31.312570 | 12.0281059 12.0281082 | 199 241 | -47 dBm -60 dBm |
| Basile | 50.024 | 30.032 | 1 hour @ 1600° C (then broke) 200° C ramp up; passive cool. | 31.340330 | 12.0267126 | ~200 | -50 dBm |

**DC-POWERED $Fe^{3+}$:sapph. MASER OSCILLATOR**

Fig. 3 and Table 2 (+ Appendix) provide a detailed anatomy and "bill of materials". The 31.3 GHz pump loop included many SMA-connectorized components (printed blue in Table 2), specified for operation over single frequency bands *below* 18 GHz. All components were screened against low-loss/spurions near 31.3 GHz prior to inclusion. The pump loop was powered by five Hittite MMIC amplifiers consuming 6.5 Watts (1.3 A at 5 V) in total.

*Mode selector:* This comprised a 6-pole (equi-ripple) 50-MHz bandwidth prefilter, (n) in Fig. 3 [retuned to 31.34 GHz; I.L. 3 dB] in series with a 3-pole, 17-MHz bandwidth filter (m) made from a 300-mm length of WR-28 waveguide; see Fig. 2. The latter was designed using [10], section 12.11; four inserted irises made from 0.2mm copper sheet, with ~2.5 mm (outer pair) and ~2.0 mm (inner pair) dia. holes, fixed with Epo-Tek H20E, formed its 3 λ/2 : 35 λ/2 : 3 λ/2 cavities.

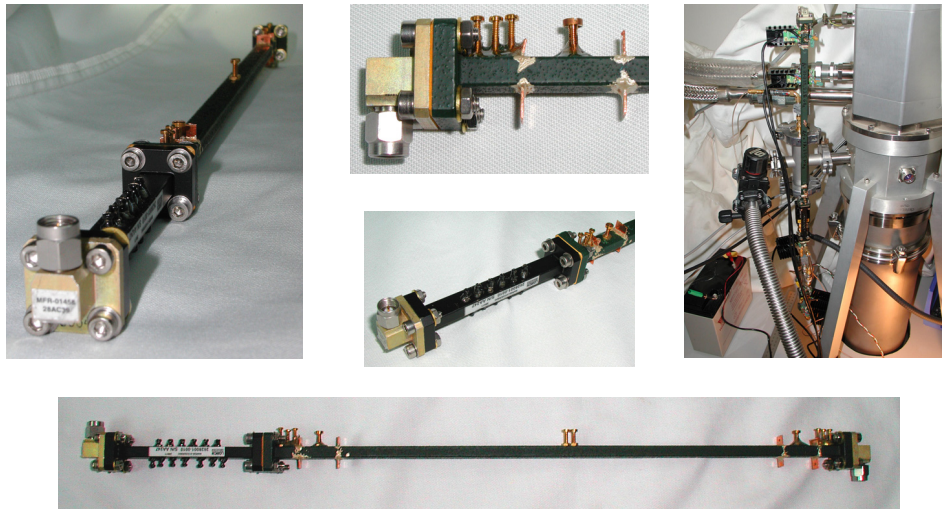

Fig. 2: WGxx mode-selection filter; top right image shows it mounted in situ.

*Voltage-controlled phase shifter:* On adjusting the manually-adjustable phase shifter (g), together with the bias voltages to each v.-c. attenuator (f), the carriers in each arm could be brought into balanced quadrature; the circuit provided 12° phase swing per volt [applied to its control circuit (y)], with ~0.3 dB/V residual a.m.; the I.L. was ~13 dB. The applied IF modulation @ 45.19 kHz gave symmetrical sidebands −15 dB down from the (recombined) carrier.

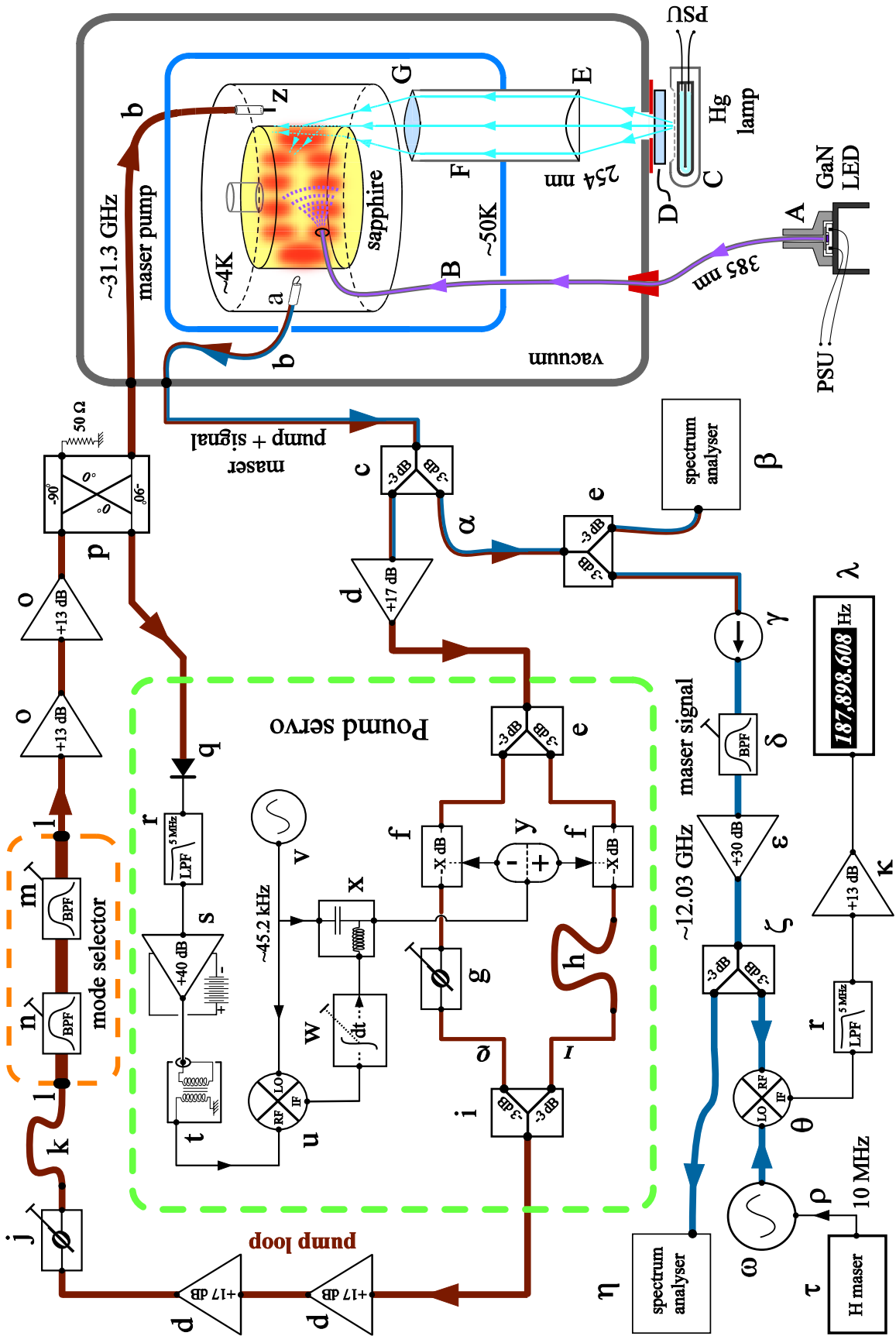

Fig. 3: Anatomy of Pound-Locked-Loop-Oscillator-Pumped Fe3+:Sapph. Maser

|   | Functional description | Make and model (+ supplier, if obscure) | Pertinent specifications and/or details (at relevant carrier frequency) |
|---|---|---|---|
|   | **Pump loop** |   | $f_P$ = ~31.3 GHz |
| a/z | loop/stub probe | lab.-crafted from RG-405U semi-rigid |   |
| b | cryo coax cable | RG178B/U, ~60cm in length | specified 1.4 dB loss per m @ 1 GHz, but … |
| c | 2-way power divider | ET Industries (Boonton NJ, USA) Model D-240-2; from *SS* | 2-40 GHz; max I.L. 2.4 dB (specified); isolation 15 dB (2-way stripline); APC 3.5 connectors |
| d | wide-band driver amplifier | Hittite HMC635LC4 mounted on eval board; | specified 18.5 dB gain; power 1.4 W (280 mA @ 5 V); noise figure 6 dB |
|   | *voltage-controlled phase shifter:* |   | overall insertion loss = -13.5 dB; phase swing from –1.0 V to +1.0 V = 24.56 degrees. |
| e | 2-way power divider | narda Model 4316.2 | 12-18 GHz nominal; but measured IL @ $f_P$ (excluding the 3dB of and ideal divider) ~1.5 dB |
| f | voltage-controlled attenuator | M/A COM Ltd; ML6550-N117-12; SPEC NO: 1085-06328; from *SS* | pair; one for each arm; 2-18 GHz; negative voltage operation; attenuation: 0 to -30 db |
|   |   |   | "*I*" channel: |
| h | fixed phase shift | looped ~8 cm length of RG-402 |   |
|   |   |   | "*Q*" channel |
| g | manually adjustable phase shifter | Spectrum C3117, LS-0212-1121; from *SS* | specification: 0-12 GHz operation; VSWR: 1.25 : 1 max; insertion loss: 0.4 db max; phase shift: 230° min; turns: 16 |
| i | 2-way power divider | narda, Model 4315.2 | specified by manufacturer for 8.0-12.4 GHz operation only; but IL = -1.2 dB @ $f_P$ measured. |
| j | manually adjustable phase shifter | MIDISCO MDC1089-1; http://www.microwavedistributors.com/; | 0-18 GHz; phase adjustment span = 10° × freq. (GHz); 0.636 ° × freq (GHz) per revolution; max length = 2 1/2 inches; measured I.L. = 0.5 dB @ $f_P$ |
| k | semi-rigid coax cable | standard RG 402 as above but ~20 cm long |   |
| l | male SMA to WR-28 waveguide adapter | MDL (Microwave Development Laboratories) 190 ELECT; MFR-01456 28AC39; from *TP* | http://www.mdllab.com/; nearest-equivalent current MDL part: 28AC226. |
| n | retuned WR-28 waveguide BPF | LORCH (http://www.lorch.com/) 6WR28-31025/R50C; from *TN* | 6-cavities (poles) in WR-28 straight, 79-mm long; originally centred on 31.025 GHz, 50 MHz bandwidth, |
| m | "mode-picker" WR-28 filter | modified 12" Flann Microwave WR28 waveguide section; from *TP* | see text for details; centre frequency 31.33522815 GHz; insertion loss 11.3 dB; -3 dB bandwidth 17.4 MHz. |
| o | medium power amp. | Hittite HMC499LC mounted on eval board; | 13 dB gain @ $f_P$ |
| p | 3 dB hybrid | MA/COM FSC 96341, PN 2032-6374-00, | 6.5–18.0 GHz; ISO port terminated by 50-Ω load |
|   | **Pound Servo** |   | $f_{IF}$ = 45,189.5 kHz; "base band"= 0-10 kHz |
| q | power detector | Agilent 8474E; negative response, 0.01-50 GHz response; 2.4-mm input connector; | ± 0.4 dB up to 26.5 GHz; sensitivity > 0.4 mV/μW to 40 GHz |
| r | low pass filter | Mini-Circuits BLP 10.7 |   |
| s | ultra-low-noise rf amplifier | lab constructed, based Plessey SL561; powered by YUASA NP17-12I rechargeable 12 V battery | 0.8 nV/√Hz input voltage noise; bandwidth 100 Hz to 6 MHz.; 40 dB gain |
| t | Balun | Mini-Circuits FTB1-1, BNC connectorized | 0.2-500 MHz |
| u | double-balanced mixer | Hatfield Instruments (Plymouth England) Modulator Type 1754 | freq. range: 0.01-100 MHz |
| v | IF frequency generator | Novotech (Seattle) Model DDS3 | 45.189.47 kHz; specified amplitude 0.7 Vpp into 50 ohms (or 0.88 dBms); 0 dBm measured. |
| w | servo loop filter = integrating amplifier | lab.-constructed; incorporating input-bias-voltage-nulled Burr-Brown OPA627. | input and output monitored with oscilloscope |
| x | passive LC bias-tee | lab.-constructed. | cross-over ~10 kHz |
| y | dual dc. bias + differential driver | lab.-constructed circuit; based around two OP177 op-amps | independently adjustable negative bias voltages. |
|   | **Maser signal chain** |   | $f_M$ = ~12.027 GHz |
| α | armoured coax cable | Midwest Microwave CSY-SSSM-52-002 MA; | 2m long |
| β | spectrum analyser | Anritsu MS2668C | 9 kHz to 40 GHz coverage; clocked by 10-MHz maser ref. |
| γ | Isolator | Aerocomm (Bangkok, Thailand) J80.124 |   |
| δ | bandpass filter | Microphase R4815, bandpass filter, centre frequency = 12.0 GHz, BW = 140 MHz; from AX | specified -0.8145 dB @ 11.94 GHz; -0.9051 dB @ 12.08 GHz; measured -0.2dB @ 12 GHz; -65 dB @ 31.3 GHz |
| ε | low-noise amp | Herotek A1242-330 | 35 dB gain; noise figure: 3.0 dB;12VDC, 350mA |
| ζ | 3dB power splitter | Advance Technical Materials, Inc. (Patchogue, NY, USA) ATM 9216 | 8.0-12.4 GHz nominal |
| ω | spectrum analyzer | HP 8566B | 2-22 GHz coverage |
| θ | Double-bal. mixer | Robinson Labs RL-7030-A-1 | 12-18 GHz nominal |
| τ | active hydrogen maser freq. reference | Sigma Tau MHM-2010; via two cascaded Spectral Dynamics, Inc. HPDA-15RM-C amplifiers | 10 MHz output; along assorted lengths of Andrews Heliax, Times Microwave LMR-400 and RG213 cables around lab. |
| ρ | 4-way passive splitter | Mini-Circuits FTB-1-1 balun into ZSC-4-1 | ~10 dBm from each ZSC-4-1 output port. |
| η | Synthesizer | Anritsu MG3692B signal generator | clocked by 10-MHz maser ref. |
| κ | rf amplifier | HP 461A | Mains powered; set to 40 dB gain |
| λ | frequency counter | Agilent 53132A | clocked by 10-MHz maser ref. |

Table 2

**Performance:** On applying current to its amplifiers, the pump loop would reliably oscillate (and Pound-lock), provided the resonator's temperature lay below 20 K. On switching off the cooler, masing would persist up to 29.5 K. The Pound detector diode (q) lay outside of the cryostat. As inferred from a germanium resistance sensor attached to the resonator's can (read with a Lakeshore 340 temperature controller), the resonator's temperature "yoyoed" with an amplitude of ~0.1 K at the cooler's cycle frequency (~1.4 Hz); a corresponding yoyoing was observed on the Pound error signal. Even when Basile was stationed at its frequency turnover temperature (8.72 K; curvature of –11.85 Hz/K$^2$) the affects of the cryocooler's thermal/mechanical vibrations still dominated; the measured Allan deviation at 10s was around 10$^{-13}$.

**BROADENING MECHANISMS**

Qualitatively, the observed weakness of the Fe$^{3+}$:sapph.'s ESRs is due to spectral (as well as spatial) hole burning. The rational design of any device based on them requires that the hole burning's dependence on variables within experimental control be quantified. This boils down to determining the values of various time constants. Where data specific to Fe$^{3+}$:sapph is lacking, estimates can be inferred from studies on (dilute) ruby; see Table 3.

*Inhomogenous (or rather "heterogenous") broadening $T_2^*$:* In the limit of low doping concentrations, Fe$^{3+}$:sapph. consistently exhibits total linewidths of a few tens of MHz, characterized by a time constant $T_2^*$ of around 10 ns; crystal quality (Verneuil, Czochralski, HEM, …) makes little difference. This intrinsic broadening is due to the hyperfine interaction between each Fe$^{3+}$ ion and the (magnetic dipole's of the) $^{27}$Al nuclei that surround it. Considering just the first 13 nearest nuclei, Wenzel predicted its magnitude of this broadening quite accurately [11]. Mössbauer spectroscopy (on $^{57}$Fe$^{3+}$-dopped sapphire) [12] subsequently clarified the earlier ESR-measurements [13]. ENDOR experiments [14, 15] motivated the concept of a "frozen core" of nuclei surrounding each Fe$^{3+}$/Cr$^{3+}$ ion.

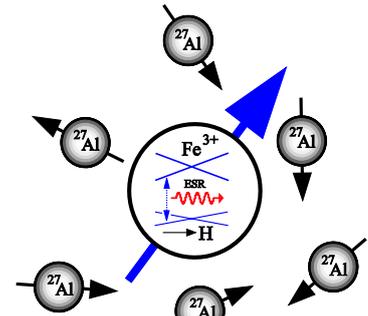
Fig. 4

*Spin-lattice relaxation $T_1$:* This parameter is also well understood [16]; it does depend on crystal quality (see Table 3.)

*Spin-spin (tranverse) relaxation $T_2$:* One problem, on the experimental side, is that most of the (surprisingly limited) relevant experimental data comes from measurements on ruby at >10 ppm concentrations. A compounding problem, on the theoretical side, is that Cr$^{3+}$:sapph at high (saturating) drive powers is known to violate Bloch theory -as based on the value of $T_2$ obtained from spin-echo ESR/ENDOR/PENDOR measurements. Even the exact shape of the homogenous broadening is still not wholly clear [17]. Acknowledging these caveats and persisting mysteries, we limit our sights to extracting a "ball-park" magnitude (for both the pump and signal transitions). The available literature indicates that, at least down to concentration levels of a few tens of ppm, the homogenous broadening of the Cr$^{3+}$:sapph. ESRs, as quantitified by $T_2$, is still controlled by Cr:Cr electronic spin flip-flops. The value of $T_2$ has been measured through a variety of techniques –see Table 3. What is essential to understand here is that $T_2$ is not simply a constant but a function of (amongst other parameters) the concentration and the drive power, *i.e.*, intensity-broadening [18] has to be included. Though violations have been observed [19], to first, rough-and-ready approximation, the homogenous broadening scales proportionally to the square-root of the concentration: $T_2 \propto 1/\sqrt{\text{conc.}}_{Fe^{3+}}$. Within Bloch theory (BT), the spreading of an excitation due to intensity broadening is quantified by [18]:

$$\Delta f = \frac{\sqrt{1+S^2}}{\pi T_2}, \qquad (1)$$

where $S = \chi\sqrt{T_1 T_2}$ is the degree of saturation, a.k.a. the dimensionless intensity; $\chi = (\gamma/2)\sqrt{\sigma\sigma^*}H$ is the Rabi frequency; $\gamma$ is the "free spin" ESR field-to-frequency conversion factor (28 GHz per Tesla); $\sqrt{\sigma\sigma^*}$ is the (magnitude of the) dipole strength for the transition concerned; H is the magnetic field amplitude.

Beyond Bloch theory, the phenomenon of spin diffusion [20] needs to be included (as through the Yamanoi-Eberly model [21] or similar), but no attempt to chart these depths will be made here. Despite addressing optical hole burning, [22] reviews what is known about spin diffusion from ESR measurements. These works study Cr$^{3+}$:sapph. in setups where the homogenous broadening grows less quickly with intensity than what BT predicts. The spin diffusion here is presumed to be driven by flip-flopping of the $^{27}$Al nuclei spins, particularly those close-lying ones in the frozen core, whose flips can "jolt" the ESR of an individual Cr$^{3+}$/Fe$^{3+}$ ion far out of resonance from its pump/single WG mode. Using nuclear spin echo in conjunction with pulsed optical excitation of ruby's R1 line, the dephasing time(s) of the $^{27}$Al : $^{27}$Al flip-flops have been measured [22, 23] for both the frozen core and bulk –see Table 3; it is worth noting here that both times are smaller than $T_1$, suggesting that a pumped $Fe^{3+}$:sapph ion will typically diffuse around the whole of the heterogenous linewidth before it non-radiatively decays. Needless to say, these measurements and insights are highly relevant to predicting the effective fraction of $Fe^{3+}$:sapph. ions that participate in maser action.

| Property | Technique | Experimental particulars | Value | Ref. |
|---|---|---|---|---|
| inhomog. linewidth = $1/\pi T_2^*$ | X-band ESR spectroscopy | 50 ppm $Fe^{3+}$<br>100-200 ppm Fe3+ | 27 ± 5 MHz<br>28 MHz | [24]<br>[25] |
| | Theoretical | for $Cr^{3+}$ (magnetic field parallel to $C$ axis) | 10 Gauss<br>= 28 MHz | |
| | Mossbauer spectroscopy | 800 ppm $^{57}Fe^{3+}$ | "9.0 ± 3.0 Gauss"<br>=25 ± 8 MHz. | [12] |
| $T_1$ | X-band ESR saturation-relaxation @ ~ 9 GHz | 50 ppm Fe3+ conc. | 7 +/- 2 ms | [26] |
| | @ 9-10 GHz | 200 ppm conc. | 8 ± 1 ms | [25] |
| | | 20 ppm conc. | 12 ± 1ms | [25] |
| | @ 9.27 GHz | residual; Verneuil (vap.-phase) | 20 ms | [27] |
| $T_2$ | echo-ENDOR; @16 and 9.3 GHz | 0.005-wt% ruby | 1.5 µs | [28] [28] |
| | echo-PENDOR@ 693.4 nm ($R_1$ line) | 0.005-wt% ruby | 3.5 µs | [29] [28] |
| | free induction decay (FID) of ESR spin @ 5.9 GHz | 0.009-wt % ruby | 7.5 ± 7 µs | [21] |
| | optical hole burning of $R_1$ line @ 693.6 nm | 0.0034-wt % ruby | 15 µs | [22] |
| $\sqrt{(T_1 T_2)}$ | microwave bistability | | 4.6 ± 0.2 µs | [30] |
| $T_2$ of $^{27}$Al | nuclear spin-echo decay | in the "frozen core" | 1 ms | [23] |
| | | in the bulk | 60 µs | ibid. |
| spin-spin (intraline cross-) relax. time | optical hole burning | 0.05 wt% ruby | 0.5 ms | [31] |
| Cr-Cr spin-flip time (reson. cross relax.) | many and various | | 1-50 µs | Table 1 in [23] |
| spectral diff. time $T_d$ | | | 14 µs | [22, 32] |

Table 3: Key design parameter for Fe3+:sapphire maser action

A final insight comes from the observation of bi-mode masering: WG modes separated by 8 MHz coexist independently [4], whereas modes stationed 10 kHz apart compete (for the same population inversion) [33] as they brighten. This suggests that the effective homogenous linewidth lies below 10 kHz at low powers and somewhere between 10 kHz and 8 MHz at high powers.

*Distilling Table 3 and the above insights into ball-park figures:* Given the low concentration levels of $Fe^{3+}$ ions in HEMEX, $T_2$ at low-power should be several tens of µs (say ~80 µs), corresponding to a homogenous linewidth of a few kHz (say ~4 kHz). With such a narrow homogenous linewidth, spin diffusion seems likely to be significant: the time, $T_d$, an $Fe^{3+}$:sapph ion typically stays within the homogenously broadened line risks being the same order of magnitude, if not shorter, than $T_2$ itself. At finite power, the key parameter is the Rabi frequency, $\chi$, as given by (1) above, where the magnetic field amplitude can be estimated through $H \approx \sqrt{QP/(\mu_0 f_0 V_{\text{eff}})}$, where $QP$, $f_0$ and $V_{\text{eff}}$ are the electromagnetic (co-circulating) power, frequency and effective volume of the WG mode in question. Compared to an X-band Ramsey cavity in a Cs clock, the extremely high $Q$s and smaller volumes of WG modes in cryogenic sapphire rings stand to elevate $\chi$ well above the tens-of-kHz Rabi frequencies encountered with the former. *Getting more quantitative (for the pump transition):* assuming an applied power of 1 mW, critical coupling, an effective mode volume of 5 cm$^3$, a (loaded) $Q$ of 1 billion; and noting that the transition amplitude of the (level-crossing thus somewhat forbidden) pump transition is only 0.05 in free spin units [26] [34], one arrives at an estimated Rabi frequency of around 1 THz; in comparison with any reasonable estimate of $1/\sqrt{(T_1 T_2)}$, this puts the saturation $S \gg 1$ corresponding to the whole inhomogenously broadened line being made accessible; Fig 6.3 and its surrounds discussion in [18] clarify these remarks. The extent to which spin-diffusion will cripple the power broadening [21], remains to be quantified.

## ULTRAVIOLET SENSITIVITY

The purpose of these experiments was to see whether either the number density or paramagnetic state of $Fe^{3+}$:sapph. could be affected by light (at lowish intensities). Many workers with interests ranging from mineralogy to gravitational wave detection have measured optical absorption spectra of sapphire specimens [35] [36]. However, because optical transitions from $Fe^{3+}$:sapph.'s electronic $^6S$ ground state are doubly forbidden, they are often masked by stronger absorptions associated with other colour centres; Refs. [37] and [38] do nevertheless make explicit identifications. In particular, they observe a relatively strong and sharp "$4T_2^b$" absorption line at 387 nm (at 77K). Although, to the best of the authors' understanding, $Fe^{3+}$:sapph is not expected (at zero applied magnetic field …) to exhibit any sort of paramagnetic circular dichroism --as would facilitate state-selective optical pumping, the extreme ease with which radiation at this wavelength (or thereabouts) can be generated with an AlN light-emitting diode motivated the shining of some of it onto the sapphire ring –just to see what might happen. This was done using a room temperature diode feeding a plastic light pipe. See Figs. 2, 5 and Table 4. A diode of the same type was also mounted on the inner wall of the copper resonator can –see right image of Fig. 5 immediately below.

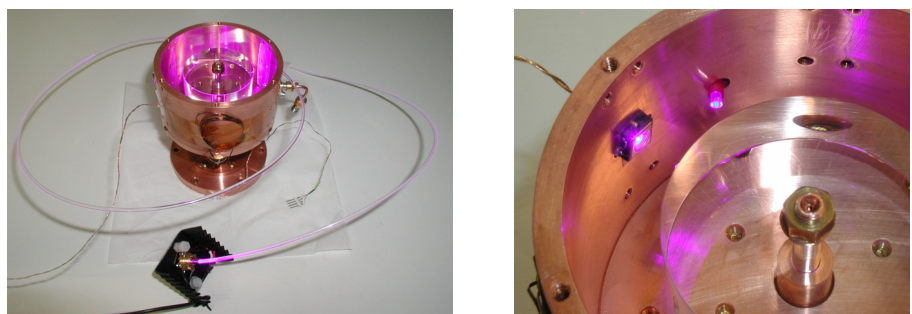

Fig.5: 385 nm optical excitation

As for the effects of even shorter-wavelength light, Tippins [39] studied the UV absorption spectrum of $Fe^{3+}$:sapph. associated with valence-changing transitions: $O^= + Fe^{3+} \rightarrow O^- + Fe^{2+}$. In particular, he found a broad absorption around 255 nm associated with adding a spin-down electron to the so-called $t_{2g}(\pi^*)$ molecular anti-bonding orbital. Optically stimulating this chemical transition should reduce, temporarily, the density of $Fe^{3+}$ ions thus the strength of their associated ESRs/masing. Again, as luck would have it, the required radiation can easily be gotten from the 254-nm line of a mercury vapour lamp. The only challenge comes in conveying it to the cryogenic sapphire ring efficiently. On removing the "lid" of the resonator can, this was done with a crystalline quartz window on the cryostat's AVC and two fused-silica convex lenses mounted within the cryostat; see Figs. 2 and 6 and Table 4.

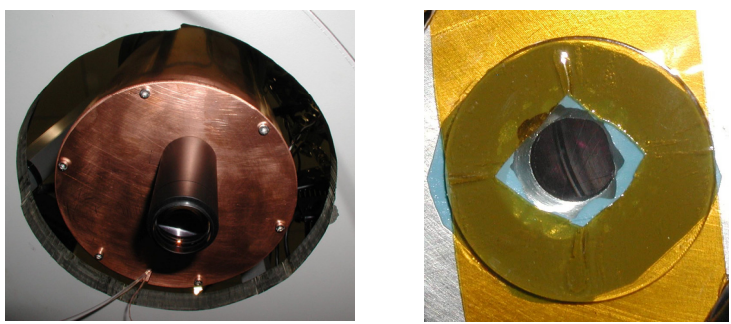

Fig 6: UV optical access of masing cryogenic sapphire ring; left: cryogenic lens assembly; right: image of cryogenic sapphire ring seen through this assembly (the circular lower and upper edges of the ring can both be seen).

**Results.** Alas, all negative! The 385 nm LED inside the copper can did not function (no light) when cold. The only noticeable effect upon injecting 385 nm radiation via the plastic light pipe was a slow transient shift in the maser frequency (taking several seconds to complete) consistent with heating. The optical power at 255 nm shone onto the sapphire crystal was estimated to be 1.63 mW (see Table 4 below). This could be turned on and off within a few tens of ms with a mechanical shutter. No substantial difference in the maser signal power could be detected when toggling between an open and closed shutter. The results of these optical-illumination experiments are, needless to say, preliminary. Using lasers or installing an optical build-up cavity would drive the excitation to $t_{2g}(\pi^*)$ harder; synchronous detection would help to detect a weak effect. It is both salutary and intriguing to note that the ESR strengths of $Cr^{3+}$:sapph. (*i.e.* ruby) were not altered upon irradiating it with sufficient $^{60}Co$ γ rays to turn it orange in colour [40].

|   | Function | Make and model | Pertinent specifications and/or details |
|---|---|---|---|
|   | **385 nm stimulation:** |   |   |
| A | 385nm UV diode | Nichia, NCSU034A | output centred on 385nm, width ~10nm; 3.8 V applied, 330 mA drawn. |
| B | Plastic light pipe | Mitsubishi "Eska" acrylic (PMMA) fibre, 2mm O.D. | 1.2 m long; attenuation @385 nm (extrapolated from measurements > 400 nm): 700 dB/km; so attenuation through pipe used ≈ 0.84 dB |
|   | **254 nm simulation:** |   |   |
| C | 255 nm mercury vapour lamp | UVP Pen-Ray lamp, model 90-0012-01 (+ 11SC-1) | housed within ~9 mm dia. aluminium shield with 0.19 ×1.5" window. |
| D | Quartz window | lab. heirloom; orientation unknown; 1" dia.; 2mm thick. | originally fine ground; hand polished against 1-micron "aloxite" held in damp cotton cloth, followed by 0.3 micron cerium oxide in same. |
| E | Fused silica lens | 1" diameter, plano-convex, focal length 50 mm nominal | manufacturer unknown; probably Spectrosil. |
| F | lens tube | Thorlabs SM1 series aluminium lens tubes of 1" dia. optics | combination of SM1L05, + SM1L20 + SM1L30 stacked together; mounted through hole in first-stage rad-shield |
| G | Fused silica lens | 1" diameter, bi-convex, focal length 50 mm nominal | manufacturer unknown; probably Spectrosil. |
|   | From http://uvp.com/mercury.html: typical intensity of Pen-Ray lamp (over the 254 nm line) at a distance of 0.75" = 4.5 mW/cm². However, "in the case of mercury lamps, the primary emissions peak at 254nm will decrease steadily with time due to quartz solarization. This will yield a net output of approximately 70% after 2,000 hours. The effect will then stabilize for the remainder of the life of the lamp." So make that 3 mW/cm². From measured experimental geometry, effective diameter of first silica lens (E) projected onto sphere 0.75"-radius sphere around from PenRay lamp ≈ 9.525 mm. Transmission losses: ~3.37% reflection loss per surface from three pairs of surfaces plus absorption through 2 mm + 6mm + 6mm of silica @ -0.5 % per mm (see http://www.uqgoptics.com/) ; compounding gives ~24 % loss in total. Thus optical power projected into sapphire ring estimated to be: π × (0.9535/2)^2 × 3 (1- 24/100) = 1.63 mW |||

Table 4

**Acknowledgements**
MO wishes to thank several NPL colleagues: Conway Langham for enlarging the cryostat's sample space; John Howes and David Gentle for loan of microwave equipment; and Roger Morrell for his assistance with the roasting.

**Appendix**

|   | Supplier |
|---|---|
| **AX** | Abex (UK) Ltd; www.abex.co.uk; tel: +44 1252 844902; email: sales@abex.co.uk; Abex (UK), Warren Close, Hartley Wintney, Hook, Hampshire, RG27 8DS, United Kingdom |
| **SS** | Solid State Electronics (UK); http://www.ssejim.co.uk/; tel: +44 2380 769598; e-mail: solidstate@ssejim.co.uk; 6 The Orchard, Bassett Green Village, Southampton, SO16 3NA, United Kingdom |
| **TP** | TestParts Sales Company; http://www.testparts.com/; 126 Railroad St., P.O. Box 425, Thomson, GA 30824 USA |
| **TN** | Telnochek, Pavel. V. Kolinko, 12610 Torrey Bluff Dr. #380, San Diego, CA, 92130, USA; e-mail: pkolinko@hotmail.com |